\PassOptionsToPackage{hyphens}{url}
\PassOptionsToPackage{usenames,dvipsnames,table}{xcolor}
\PassOptionsToPackage{breaklinks,colorlinks}{hyperref}
\documentclass[sigplan,10pt]{acmart}

\usepackage{amsmath,amsopn}
\usepackage{float}
\usepackage{subfigure}
\usepackage{endnotes,microtype,xspace,graphicx,fancyvrb,multirow}
\usepackage{booktabs}
\usepackage{array,underscore,relsize}
\usepackage[T1]{fontenc}
\usepackage{fancyhdr,totpages}

\usepackage{enumitem}
\usepackage[labelfont=bf,font=small,skip=5pt]{caption}
\usepackage{fp}
\usepackage{siunitx}

\usepackage{balance}

\usepackage{verbatim}

\usepackage{listings}
\usepackage{color}
\definecolor{lightgray}{rgb}{0.92, 0.92, 0.92}
\definecolor{darkgray}{rgb}{0.4, 0.4, 0.4}
\definecolor{purple}{rgb}{0.65, 0.12, 0.82}
\definecolor{editorGray}{rgb}{0.95, 0.95, 0.95}
\definecolor{editorOcher}{rgb}{1, 0.5, 0} 
\definecolor{editorGreen}{rgb}{0, 0.5, 0} 
\definecolor{codegreen}{rgb}{0,0.6,0}
\definecolor{darkgreen}{rgb}{0,0.45,0}
\definecolor{codeblue}{rgb}{0,0,0.55}
\definecolor{linkblue}{rgb}{0,0,0.8}
\definecolor{darkred}{rgb}{0.8,0.1,0.15}
\definecolor{codegray}{rgb}{0.5,0.5,0.5}
\definecolor{codepurple}{rgb}{0.58,0,0.82}
\definecolor{backcolour}{rgb}{0.95,0.95,0.92}
\definecolor{mauve}{rgb}{0.58,0,0.82}

\usepackage[ruled, vlined, inoutnumbered, linesnumbered]{algorithm2e}
\usepackage{algorithmic}

\newcommand{\sys}{\mbox{\textsc{R1-Fuzz}}\xspace}
\makeatletter
\patchcmd{\lsthk@TextStyle}{\let\lst@DefEsc\@empty}{}{}{\errmessage{failed to patch}}
\makeatother

\newcommand{\TODO}[1]{\textcolor{Melon}{TODO: #1}}

\newenvironment{mybullet}{\begin{list}{$\bullet$}
		{\setlength{\topsep}{0.5mm}\setlength{\itemsep}{0.5mm}
			\setlength{\parsep}{0.5mm}
			\setlength{\itemindent}{0.5mm}\setlength{\partopsep}{0.5mm}
			\setlength{\labelwidth}{15mm}
			\setlength{\leftmargin}{4mm}}}{\end{list}}
\newcommand*\circled[1]{\tikz[baseline=(char.base)]{
            \node[shape=circle,draw,inner sep=0.5pt] (char) {#1};}}

\fvset{fontsize=\scriptsize,xleftmargin=8pt,numbers=left,numbersep=5pt}

\input{code/fmt}

\def\Snospace~{\S{}}

\newif\ifdraft\drafttrue
\newif\ifnotes\notestrue
\ifdraft\else\notesfalse\fi

\newcommand{\eg}{{\em e.g.}}

\newcommand{\ie}{{\em i.e.}}

\input{glyphtounicode}
\newcolumntype{R}[1]{>{\raggedleft\let\newline\\\arraybackslash\hspace{0pt}}p{#1}}

\newcommand{\squishlist}{
\begin{itemize}[noitemsep,nolistsep]
  \setlength{\itemsep}{-0pt}
}
\newcommand{\squishend}{
  \end{itemize}
}

\usepackage{tikz}

\usepackage{xstring}
\newcommand{\PP}[1]{
\vspace{2px}
\noindent{\bf \IfEndWith{#1}{.}{#1}{#1.}}
}

\newcommand{\boxbeg}{
\vspace{2px}
\noindent\begin{tabular}{|l|}\hline
\begin{minipage}{3.2in}
\vspace{2px}
\noindent
}

\newcommand{\boxend}{
\vspace{2px}
\end{minipage}\\ \hline
\end{tabular}
\vspace{-10pt}
}

\ifdefined\DRAFT
\fi

\author{Jiayi Lin}
\affiliation{%
    \institution{The University of Hong Kong}
    \city{Hong Kong SAR}
    \country{China}
}
\email{linjy01@connect.hku.hk}

\author{Liangcai Su}
\affiliation{%
    \institution{The University of Hong Kong}
    \city{Hong Kong SAR}
    \country{China}
}
\email{liangcaisu@connect.hku.hk}

\author{Junzhe Li}
\affiliation{%
    \institution{The University of Hong Kong}
    \city{Hong Kong SAR}
    \country{China}
}
\email{jzzzli@connect.hku.hk}

\author{Chenxiong Qian} \authornote{Corresponding author.}
\affiliation{%
    \institution{The University of Hong Kong}
    \city{Hong Kong SAR}
    \country{China}
}
\email{cqian@cs.hku.hk}

\setcopyright{none}
\renewcommand\footnotetextcopyrightpermission[1]{} 

\title{\sys: Specializing Language Models for Textual Fuzzing via Reinforcement Learning}

\begin{document}
\begin{abstract}

Fuzzing is effective for vulnerability discovery but struggles with complex targets such as compilers, interpreters, and database engines, which accept textual input that must satisfy intricate syntactic and semantic constraints. 
Although language models (LMs) have attracted interest for this task due to their vast latent knowledge and reasoning potential, their practical adoption has been limited.
The major challenges stem from insufficient exploration of deep program logic among real-world codebases, and the high cost of leveraging larger models.
To overcome these challenges, we propose \sys, the first framework that leverages reinforcement learning (RL) to specialize cost-efficient LMs and integrate them for complex textual fuzzing input generation.
\sys introduces two key designs: \textit{coverage-slicing-based} question construction and a \textit{distance-based} reward calculation. 
Through RL-based post-training of a model with our constructed dataset, \sys designs a fuzzing workflow that tightly integrates LMs to reason deep program semantics during fuzzing.
Evaluations on diverse real-world targets show that our design enables a small model, named \sys-7B, to rival or even outperform much larger models in real-world fuzzing.
Notably, \sys achieves up to 75\% higher coverage than state-of-the-art fuzzers and discovers 29 previously unknown vulnerabilities, demonstrating its practicality.


\end{abstract}

\maketitle

\sloppy
\section{Introduction}
\label{s:intro}


Software fuzz testing, or \textit{fuzzing}, is an effective technique for uncovering hidden security vulnerabilities by generating diverse inputs to execute the programs under testing, aiming at exploring deep semantic logic \cite{fuzzbench, ossfuzz, libfuzzer}.
One of the most persistent challenges in fuzzing lies in the input generation for target programs that handle complex structured inputs, such as language compilers, interpreters, databases, etc. 
These targets expect inputs that conform to intricate grammars, embody rich semantic rules, and engage deep program logic. 
Existing approaches \cite{nautilus, gramatron, grammar, polyglot} usually suffer from scalability and adaptability issues.
Besides requiring non-trivial manual labor for each target, any evolution of the target program, such as new language features, might render their handcrafted specifications, mutators, or ad-hoc rules incomplete, leaving program states unexplored.
In this paper, we refer to fuzzing such targets as \textit{textual} fuzzing, where the input complexity arises from structured printable formats, and focuses on the central challenge of \textit{generating high-quality fuzzing inputs for complex textual targets}.

For textual fuzzing, large language models (LLMs) offer transformative potential for overcoming these limitations. 
First, they possess vast latent knowledge from pretraining on large programming and natural language datasets, equipping LLMs with a "universal corpus" that encodes syntax, semantics, and idioms of the targets under testing.
Second, LLMs exhibit strong semantic understanding ability. 
They can reason about program intent from rich semantic cues, not only from structural code, but also natural language semantics inside code symbol names or comments, which are extremely challenging for conventional methods to harness. 
This allows LLMs to infer how to construct meaningful, syntactically correct, and semantically relevant inputs that reach deep program branches. 
For instance, \autoref{code:example1} is an example we observed during our experiments.
In the target program \textit{SQLite}, the conditional branch after line 9 is consistently unreached by traditional fuzzers, while LLM can easily generate relevant inputs (line 13) to satisfy the condition, improving fuzzing code coverage. 
It is because the code comments in lines 4 to 8 provide functionality explanations of the following branch (SQL queries with \lstinline{RANGE BWTWEEN a FOLLOWING AND b FOLLOWING} where \lstinline{a} equals \lstinline{b}) that LLM can understand and harness.

{
\begin{lstlisting}[
    caption={Example of pervasive, underutilized semantic cues (\eg{,} code comments, symbol names) that are suitable for language models to harness during fuzzing input generation, from \textit{SQLite}.},
    label={code:example1},
    float=!t,
]
// sqlite/src/window.c
static int windowCodeOp(WindowCodeArg *p, ...
  ...
/* If this is a (RANGE BETWEEN a FOLLOWING AND b 
** FOLLOWING) or (RANGE BETWEEN b PRECEDING AND 
** a PRECEDING) frame, ensure the start cursor 
** does not advance past the end cursor within the
** temporary table. ... */
  if( pMWin->eStart==pMWin->eEnd && regCountdown
      && pMWin->eFrmType==TK_RANGE ...

// A part of test input successfully generated by LLMs (by understanding the comments above) to satisfy and test the conditional branch:
SELECT b OVER w FROM t WINDOW w AS (ORDER BY b RANGE BETWEEN 10 AND 10);
\end{lstlisting}

However, the reasoning potential of LLMs has not been fully leveraged in practice.
A primary obstacle is the challenge of applying them effectively to large-scale, real-world codebases. 
Although such codebases contain rich semantic information crafted for human comprehension, their scale exacerbates well-known limitations of LLMs: the trade-off between hallucination and context window sizes \cite{hallucination}. 
Moreover, the semantic cues required to reach specific program branches are often fragmented throughout deep or nested call chains, making them difficult to isolate and present meaningfully to LLMs.
Therefore, constructing targeted prompts that reliably guide LLMs to explore deep program logic for fuzzing remains unaddressed.
As a result, existing approaches often circumvent this issue by leveraging LLMs on curated, targeted documentation \cite{fuzz4all, meng2024large} or isolated fuzzing inputs \cite{covrlfuzz, clozemaster} that have a manageable size. 
To the best of our knowledge, there is no existing work that harnesses LLMs for fuzzing by directly reasoning over the target codebases.
To summarize, the crucial challenge lies in: \textit{constructing reasonable questions from large real-world code bases to explore deep program logic (\textbf{C1})}.

Furthermore, applying LLMs inherently faces the \textit{cost and scalability concerns (\textbf{C2})}.
Existing strategies generally rely on querying larger models for better performance.
However, large models are expensive to query repeatedly, especially within intensive tasks like fuzzing input generation. 
As a result, some works choose to offload LLMs outside the fuzzing loop, such as generating mutator code \cite{metamut, zhang2025low} or driver code \cite{promptfuzz, rug} offline.

Fortunately, recent research has introduced a promising paradigm that employs RL-based post-training \cite{deepseekr1} to align the reasoning capabilities of smaller models, enabling them to specialize in domain-specific tasks \cite{compilerr1, swerl, finr1, memoryr1, rlsf}.
Such an RL framework for LMs requires both well-structured questions to elicit high-quality answers and a goal-aligned reward mechanism to evaluate those answers and guide the models' parameter updates.
However, existing reward strategies for LMs rely on textual pattern matching \cite{logicr1} or simplistic execution output comparisons \cite{codeio}, which are inaccurate or too sparse to effectively guide fuzzing input generation, presenting a key challenge of \textit{designing a precise and informative reward signal (\textbf{C3})}.

To systematically overcome these challenges, we present \sys, a novel framework for textual fuzzing powered by LLMs with reinforcement learning-based post-training.
\sys first trains a cost-efficient model to specialize at the task of textual input generation (addressing \textbf{C2}) with a novel reward signal, and then uses it to generate high-quality inputs in real-world fuzzing campaigns.
Specifically, \sys introduces two key designs: \textbf{coverage-slicing-based question construction} and \textbf{distance-based reward calculation}.
To address \textbf{C1}, \sys decomposes the program by executing a seed input to identify nearby uncovered branches (\ie{,} conditional paths not taken). 
For each target branch, it constructs a focused question by slicing the source code along the execution path from the program entry to that branch.
This question construction method is used in both generating datasets for training and generating questions on the fly during fuzzing.
Next, to address \textbf{C3}, our distance-based reward mechanism provides a finer-grained signal by computing the distance between the execution path of an LM-generated input and the target branch. 
This yields a more accurate and less sparse training signal, inspired by research on directed fuzzing \cite{directed}.

Based on these two designs, \sys consists of three stages.
{\circled{1} Dataset Construction}: using our coverage-slicing-based question construction, given a target program and an input seed, \sys can generate a set of questions.
For a given set of seeds (\ie{,} corpus), \sys can generate a variety of questions leading to diverse deep code paths of the target program.
Furthermore, by incorporating more target programs, \sys can construct a comprehensive dataset consisting of diverse real-world source code that captures a wide range of program semantics.
The second stage is: {\circled{2} RL-based Post-training}. \sys uses the Group Relative Policy Optimization (GRPO) algorithm \cite{deepseekr1} to train a language model to excel at our task.
By integrating our reward mechanism, \sys stimulates the potential of a cost-efficient small model, named \textbf{\sys-7B}, to effectively generate targeted fuzzing inputs.

The third stage in \sys is a: {\circled{3} LLM-powered Fuzzing Loop}.
We design a fuzzing workflow that can tightly integrate LMs into a coverage-guided fuzzing loop.
%
Specifically, the loop maintains an evolving fuzzing corpus of input seeds that trigger new code coverage, and \sys uses these seeds to continuously construct questions, which express uncovered branches by the fuzzer.
During fuzzing, the model responds to the questions with new seeds that it predicts will reach a specified uncovered branch.
The new seeds then enrich the fuzzing corpus for further mutation by the fuzzer, forming a loop where our model exerts its potential to reason deep program logic guided by code coverage, eventually facilitating vulnerability discovery.
Moreover, beyond the generated static dataset, our fuzzing loop establishes a benchmark, enabling the integration and practical evaluation of various LMs to assess their real-world fuzzing capabilities.

To evaluate our design, we implemented \sys and conducted comprehensive experiments to assess both the training and real-world fuzzing effectiveness. 
We selected a diverse set of \textit{textual} targets, including language interpreters (PHP, Lua, Ruby, QuickJS, NJS), compilers (CPython, Solidity), and database engines (SQLite, DuckDB). 
We then performed all three stages in \sys on these targets, including generating a dataset consisting of 16338 questions, training a low-cost model (\sys-7B) based on Qwen2.5-7B-Instruct \cite{qwen}, and comparing the performance across different models and state-of-the-art fuzzers.
Moreover, we also conducted an ablation study to demonstrate the effectiveness of the key components in \sys.
Our experiment results showed that \sys-7B can excel at the task of fuzzing input generation, comparable to or even outperforming larger and costly models (DeepSeek-V3 \cite{deepseekr1}, GPT-o4mini \cite{o4mini}) in real-world fuzzing.
Furthermore, our LLM-powered fuzzing loop showed impressive performance on both code coverage and vulnerability discovery.
On our selected targets, \sys achieved 75\% higher coverage than state-of-the-art fuzzers \cite{nautilus, gramatron, polyglot, aflpp, libfuzzer} and discovered 29 previously unknown vulnerabilities, with 24 fixed or confirmed by the developers.
Lastly, the ablation study showed that our question format and scheduling outperforms naive strategies by 8\% and 3.8\% on correct answer ratio and code coverage.

In conclusion, we made the following contributions:
\begin{mybullet}
\item We designed a \textbf{coverage-slicing-based question construction} technique to decompose large codebases into targeted prompts, forming a dataset for LMs to reason about deep program semantics for fuzzing.
\item We designed a \textbf{distance-based reward mechanism} that provides a fine-grained training signal, enabling efficient RL-based post-training of a low-cost model (\sys-7B) to specialize in fuzzing input generation.
\item We implemented a novel coverage-guided \textbf{LLM-powered fuzzing loop}. It achieves a 75\% higher code coverage and discovers 5x more previously unknown bugs compared to state-of-the-art fuzzers.
\item We open-sourced our model and implementation, named \sys (\url{https://github.com/HKU-System-Security-Lab/R1-Fuzz}), as a comprehensive framework to post-train, deploy, and evaluate different LMs on real-world fuzzing tasks.
\end{mybullet}

\section{BackGround and Motivation}
\label{sec:bg}

\subsection{Fuzzing \textit{Textual} Targets}


Many software systems process complex structured text, such as compilers, interpreters, and database engines. These systems employ multi-phase pipelines where inputs (e.g., source code, SQL queries) must first be syntactically valid and then semantically meaningful to exercise deep execution logic. Fuzzing these textual targets is uniquely challenging: to trigger deep behaviors, inputs must satisfy intricate grammar rules and semantic constraints (e.g., type systems, variable scoping). Furthermore, these targets evolve rapidly with new language features and dialects, rendering static grammar-based testing setups obsolete. We define such systems, whose input complexity stems from printable text, as \textit{textual} targets.

Traditional approaches to test textual targets include:
\textit{Grammar-aware mutators} (e.g., Nautilus \cite{nautilus}, Gramatron \cite{gramatron}) preserve syntactic validity but might neglect semantic correctness, such as producing undefined variables or type errors that cause early rejection during execution.
\textit{Handcrafted specifications} encode both syntax and semantic rules (e.g., Polyglot \cite{polyglot} uses ANTLR/Flex grammars with additional ad-hoc rules). 
However, writing and maintaining such specifications is labor-intensive and often fails to fully capture all behaviors of complex targets.
\textit{Manually curated seed corpora} (e.g., \cite{polyglot, google-fuzzilli}) provide domain-specific seeds to guide mutation, but their effectiveness is limited by the scope and quality of available seeds, leaving many feature-specific code paths untested.

To summarize, the crucial challenge for fuzzing complex textual targets persists: \textit{testing textual targets involves deep semantic logic layered on top of complex syntactic rules, making high-quality input generation difficult and fragile.}
This challenge motivates our exploration of adopting LLMs, which are inherently equipped with latent knowledge of programming languages and structures, offering the potential to bridge both syntax and semantics in a generalizable way. 
We discuss this potential in the following subsection.

\subsection{Motivation Examples}

{
\begin{lstlisting}[
    caption={An example of LLM utilizing semantic cues from code comments and symbol names to infer relevant input, reaching a deep program branch that hinders traditional testing tools.},
    label={code:example2},
    float=!t,
]
// solidity/libsolidity/ast/Types.cpp
BoolResult StructType::isImplicitlyConvertibleTo(Type const& _convertTo) const
{
  if (_convertTo.category() != category())
    return false;
  auto& convertTo = dynamic_cast<StructType const&>(_convertTo);
  // memory/calldata to storage can be converted, but only to a direct storage reference
  if (convertTo.location() == DataLocation::Storage && location() != DataLocation::Storage && convertTo.isPointer())
  ...

// Correctly generated input from our language model to satisfy this conditional branch
contract test {
  struct S { uint x; }
  S public foo;
  function test() public {
    S memory bar = S({x: 1});
    foo = bar; // convert data type located in memory to storage
  }
}
\end{lstlisting}

To illustrate the unique capability of LLMs in textual fuzzing, consider the example shown in \autoref{code:example2}.
It comes from the compiler for Solidity, a smart contract programming language.
The function \lstinline{isImplicitlyConvertibleTo} started from line 4 checks if a \lstinline{struct} type can be implicitly cast to another \lstinline{_converTo} type.
The conditional branch (line 10) during fuzzing in our experiments was consistently left unexplored in our experiment with state-of-the-art fuzzers, while our LLM-powered fuzzing loop successfully generated an input (lines 14 to 21) to satisfy it and improved the code coverage.
The crucial part is constructing a new variable \lstinline{foo} resided in the public \lstinline{Storage} location (line 15, similar to a global variable) and assigning to it with a variable (\lstinline{bar}) resided in the \lstinline{Memory} location (lines 18 to 19, similar to a local variable), triggering an implicit conversion.
This is challenging for traditional mutation-based fuzzers because it requires multiple coordinated steps (defining variables, ensuring type consistency, performing the assignment).
Moreover, this function invocation is deeply nested in the execution call stack (with a depth of 21), making testing techniques like symbolic or concolic execution easily fail due to path or state explosion \cite{sesurvey}.

In contrast, this example is intuitive for humans with a background knowledge of Solidity, who only need to take the hints from the comment in line 9.
Similarly, LLMs can exploit such explicit cues, leveraging their pretrained knowledge of programming languages. 
We also performed an experiment that removes the comment in line 9.
The result showed that LLMs can still generate the desired input by inferring the intended behavior from \textit{symbol names} in the source code (\eg{,} \lstinline{ConvertbleTo}, \lstinline{Storage}, \lstinline{isPointer}, etc).
Notably, such semantic cues are pervasive across complex software systems because they are written for human readability.
Therefore, equipped with human-like capabilities of semantic interpretation, LLMs are uniquely positioned to harness these signals for the task of fuzzing input generation.

However, despite its potential, directly leveraging an LLM in practical fuzzing faces the following challenges:
\textit{\textbf{C1:} Constructing reasonable questions from large real-world codebases to explore deep program logic.}
Large programs contain thousands of branches, and it is non-trivial to extract relevant prompts that highlight specific code logic for the LLM to reason about. 
To the best of our knowledge, there is no existing work that harnesses LLMs for fuzzing by reasoning over the target codebases to uncover deep program logic.
Existing methods often employ LLMs on carefully selected documentation \cite{fuzz4all, meng2024large} or isolated fuzzing inputs \cite{covrlfuzz, clozemaster}, which are easier to manage in size.
\textit{\textbf{C2}: Managing inference cost and scalability}.
Larger LMs generally perform better, but are inevitably expensive to use in the intensive fuzzing loop. 
Some existing works \cite{metamut, zhang2025low} choose to offload LLMs outside the fuzzing loop, such as generating mutator code offline, to avoid querying costly LLMs intensively.

In this paper, we address \textit{\textbf{C1}} by designing a coverage-slicing technique that decomposes programs into compact, branch-specific prompts. 
This technique is lightweight enough to be conducted during fuzzing and, more importantly, can be used to construct a training dataset as illustrated in the next subsection.
To address \textit{\textbf{C2}}, we design a reinforcement learning-based post-training framework for smaller LMs, which can make them achieve comparable or even better performance than larger ones, significantly reducing deployment cost. 
Next, we discuss our motivations for RL-based post-training and the associated challenges.

\subsection{Reinforcement Learning-based Post-training}

In typical reinforcement learning (RL) for LLM setups, the model receives a prompt (\ie{,} \textit{question}), generates a response, and receives a reward that reflects how well the response satisfies the objective. 
The reward is then used to adjust the model’s parameters via policy gradient methods such as PPO or GRPO \cite{deepseekr1}.
Previous efforts \cite{swerl, compilerr1, rlsf, finr1, memoryr1, logicr1} have shown that RL-based post-training smaller models can significantly improve performance, making them competitive with much larger models at a fraction of the cost.
Inspired by them, we aim to use RL to train LLMs for fuzzing input generation.

However, existing works depend heavily on their domain-specific dataset and reward signal, which can not be directly applied to our task.
Therefore, the first critical component is \textit{question dataset construction for textual fuzzing}.
To train the model effectively, we must construct a dataset of questions that aligns with the fuzzing objective, \ie{,} generating inputs that explore designated uncovered code regions. 
To the best of our knowledge, there is no existing dataset for this purpose. 
This challenge aligns with \textbf{\textit{C1}} described above, because it is natural to standardize the question format \textit{for both model training and its application in actual fuzzing}, ensuring consistency.

The second critical component in RL is the reward mechanism.
Most existing LLM RL work uses reward signals such as string pattern matching \cite{deepseekr1, logicr1} or execution output comparisons \cite{codeio}. 
These are inadequate in fuzzing, where the objective is not to match a fixed output but to maximize the exploration of diverse program states.
Therefore, we need a reward that reflects how “close” an input is to reaching a desired program location,
which leads to the challenge \textit{\textbf{C3}: designing a precise, non-sparse reward function that guides the model training}.

To summarize, fuzzing textual targets presents unique challenges due to deep syntax and semantic constraints.
Using LLMs further faces the need for semantically relevant prompts (\textbf{\textit{C1}}) and the practical cost limitations of using large models during fuzzing (\textbf{\textit{C2}}). 
While reinforcement learning offers a promising path to address \textbf{\textit{C2}}, it introduces a new challenge in designing an effective reward mechanism (\textbf{\textit{C3}}).
These challenges motivate our design of \sys in this paper, a framework that addresses these issues through coverage-slicing-based question construction (\autoref{sec:design:dataset}) and RL-based post-training (\autoref{sec:design:train}).
Finally, \sys harnesses our specialized, low-cost model in an LLM-powered fuzzing loop (\autoref{sec:design:fuzz}) to facilitate real-world testing and vulnerability discovery.


\section{\sys Design}
\label{sec:design}

\subsection{Overview}

\begin{figure*}
    \centering
    \includegraphics[width=0.9\textwidth]{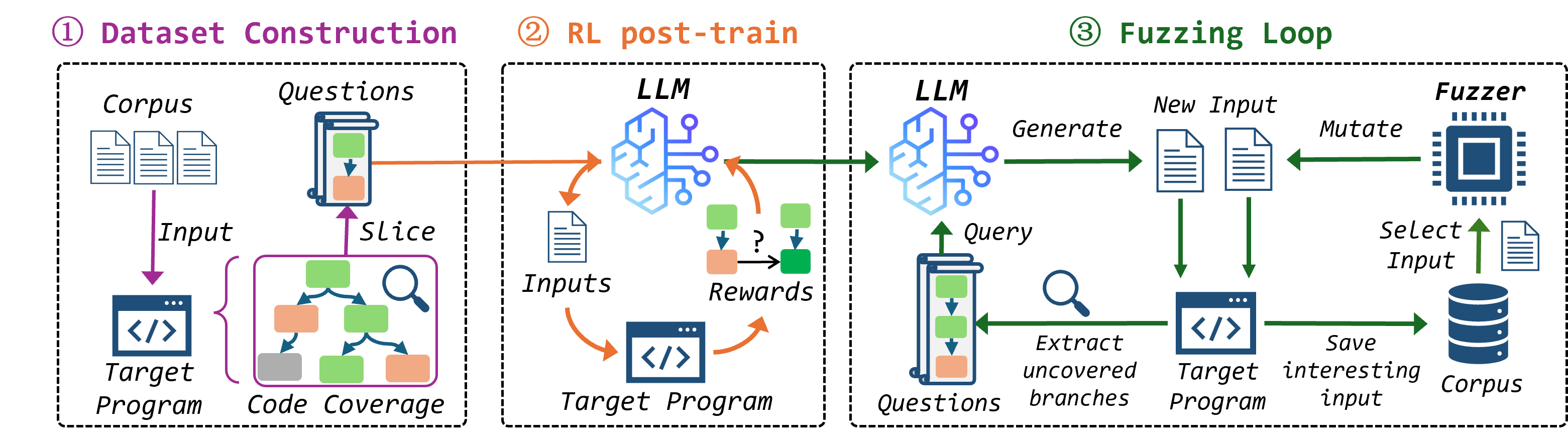}
    \caption{Overview of \sys.}
    \label{fig:overview}
\end{figure*}

\autoref{fig:overview} illustrates the overview of our \sys framework. 
It trains language models (LMs) to excel at textual input generation for fuzzing and vulnerability discovery. 

The first stage of \sys is \textit{Dataset Construction} (stage \circled{1}, \autoref{sec:design:dataset}). 
Reinforcement learning (RL)-based post-training typically requires a high-quality dataset that is task-specific and informative.
To address this, we design an automatic method to construct a dataset from scratch. 
The dataset consists of a set of \textit{questions}, which embody relevant source code for the model to reason about.
To construct them, \sys begins by collecting an initial input corpus for each target program.
By executing the target with inputs, stage \circled{1} performs code slicing based on run-time coverage to extract multiple \textit{code traces}, \ie{,} the source code alongside the execution paths. 
Each code trace starts from the program entry and ends at one branch condition (e.g., \lstinline{if} or \lstinline{case} inside a \lstinline{switch}) that an input encountered during execution.
We extract branches that are \textit{uncovered} by the given input, specifically those where the condition is evaluated to one outcome (e.g., False) while the alternative outcome (e.g., True) remains unexplored.
We then form a question by presenting the code trace and the original input to the LM with the instruction to \textit{generate a new input that inverts the evaluation outcome of this branch condition (False to True)"}. 
The LM's objective is thus to generate inputs that flip the branch decision, guiding the exploration into uncovered code regions.

Following dataset construction, stage \circled{2} is post-training via RL, in which the LM repeatedly answers our questions and receives reward feedback (\autoref{sec:design:train}).
An effective reward mechanism is critical for producing goal-aligned responses.
To address this, we design a new mechanism named \textit{distance-based reward} (\autoref{sec:design:train}), which executes LM-generated inputs and computes the reward based on how closely their execution paths approach the specified target branch.

In stage \circled{3}, a post-trained LM is integrated into a fuzzing loop for practical evaluation (\autoref{sec:design:fuzz}).
As shown in the figure, the right half of the loop follows a traditional workflow: the fuzzer mutates seeds from a corpus, executes the target, and retains inputs that discover new coverage to evolve the corpus.
To integrate the LM, the left half augments this process. 
\sys analyzes the corpus to identify branches that remain uncovered by the fuzzer, and then uses coverage slicing to generate questions from these branches to query the model.
By responding to these questions during fuzzing, the model generates high-quality inputs that target specific unexplored regions and also enrich the corpus for further mutation. 


\subsection{Dataset Construction}\label{sec:design:dataset}

The dataset of \sys consists of a set of questions for RL-based post-training. 
In textual fuzzing, the targets usually feature large codebases with numerous branch conditions to process complex textual inputs.
Therefore, it is crucial to determine a concise and effective question format to exert LMs' reasoning ability.
In \sys, we mimic how humans debug and reason about program semantics: using a concrete input to trace execution paths and reason about specific program locations.
Inspired by this, we develop a \textbf{coverage-slicing-based question construction} method.
Each constructed question contains a sliced source code trace, a specific branch position, the content of the original input, and instruction prompts.
The targeted branch is guarded by a conditional statement whose opposite outcome was left unexplored by the original input. 
The model's objective is to generate new inputs that satisfy the unexplored condition (e.g., invert \lstinline{False} to \lstinline{True}), thereby aligning its reasoning with the fuzzing objective of covering untested code.



\begin{algorithm}[htbp]
\caption{Datsset Construction}
\label{alg:dataset}
\SetKwProg{Fn}{Function}{:}{end}
\SetKwFunction{FConstructQuestions}{ConstructQuestion}
\SetKwFunction{FConstructDataset}{ConstructDataset}

\Fn{\FConstructQuestions{$\mathit{Branch}$}}{
    $\mathit{functions} \gets \mathsf{ExtractSlice}(\mathit{Branch.function})$\;
    $\mathit{question.prompt} \gets \mathit{SYSTEM\_PROMPT}$
    $\quad \| \, \mathsf{Concat}(\mathit{functions}) \, \| \, \mathit{SUFFIX\_PROMPT}$\;
    $\mathit{question.branch} \gets \mathit{Branch}$\;
    \Return $\mathit{question}$\;
}

\vspace{0.5em}
\Fn{\FConstructDataset{$\mathit{Program}$, $\mathit{Seeds}$}}{
    $\mathit{dataset},\mathit{Coverage},\mathit{BranchSet} \gets \varnothing$\;
    \ForEach{$\mathit{seed} \in \mathit{Seeds}$}{
        $\mathit{feedback} \gets \mathsf{Execute}(\mathit{Program}, \mathit{seed})$\;
        $\mathit{Coverage} \gets \mathit{Coverage} \cup \mathit{feedback.CoveredBranches}$\;
        \ForEach{$\mathit{branch} \in \mathit{feedback.UncoveredBranches}$}{
            $\mathit{BranchSet}.\mathsf{Add}(\mathit{branch})$\;
        }
    }
    \ForEach{$\mathit{branch} \in \mathit{BranchSet}$}{
        \If{$\mathit{branch} \in \mathit{Coverage}$}{
            $\mathit{question} \gets \FConstructQuestions(\mathit{branch})$\;
            $\mathit{dataset}.\mathsf{append}(\mathit{question})$\;
        }
    }
    
    \Return $\mathit{dataset}$\;
}

\end{algorithm}

\autoref{alg:dataset} describes the process of dataset construction.
The \lstinline{ConstructDataset} function first receives the target program and an initial set of input seeds (\ie{,} corpus) as arguments.
Then, the construction begins by executing the program with each seed and collecting the accumulated code coverage (lines 8 to 12).
For each seed, its uncovered branches are uniquely identified by its source code location and stored in the \lstinline{BranchSet}, meaning the execution of this seed never reaches them.
Next, after the execution of all seeds, the algorithm examines each uncovered branch to check if it belongs to the \lstinline{Coverage} set, which means that \textit{there exist other seeds that reach this branch} (line 14).
This ensures that all constructed questions have answers, \ie{,} exists a feasible input to satisfy the condition (not dead code).
While most branches in real-world targets are live, our evaluation uses targets with human-written fuzzing drivers, which may apply option settings that render certain code regions unreachable.
Therefore, we apply this filtering strategy, which is also configurable to allow a trade-off between question validity and dataset size.
The seeds that cover a branch serve as ground-truth answers to its question.
Although they are not used in RL (stage \circled{2}), which relies on reward signals rather than ground truth labels, the answer seeds are supplied into the initial fuzzing corpus to prevent data leakage and guarantee fair evaluation of generalization (\autoref{sec:eval:fuzz}).

Next, the \lstinline{ConstructQuestion} function generates one question given one uncovered branch (lines 1 to 5).
It first extracts the runtime call stack to identify functions leading to the branch. 
To provide accurate context, we statically parse and analyze the program to maintain context information, including call graphs, function bodies, call site locations, etc
Then it forms the question body by concatenating instruction prompts with extracted code.
The system prompt instructs the LMs with the objective description and format requirement for extracting generated seeds.
The suffix prompt, or user prompt, specifies the target branch condition and the original and desired outcome: \textit{Generate a new input to invert the branch condition \lstinline{COND}'s outcome from \lstinline{False/True} to \lstinline{True/False}}.
Besides the prompts and code, we also provide the original input content in the question, which is omitted in the algorithm for brevity.


\PP{Question Validity}
Our design of the question format aims to balance reasonable lengths and context validity (or completeness), \ie{,} \textit{not all} necessary context is provided in the question body.
For example, executed and returned functions not in the current call stack might also be necessary or even more helpful to reason about the target branch condition.
Such a trade-off is necessitated by LMs' token limits, the scale of real-world codebases, and the efficiency demands of RL training.
Nevertheless, given that LLMs are pretrained on vast and diverse corpora, their latent knowledge can also compensate for the missing context and enable effective reasoning.
Our experiments (\autoref{sec:eval:abl}) explored the impact of question formats and demonstrated that our design can provide effective context for LMs to facilitate fuzzing performance in real-world testing.
We left the exploration of more effective question formats as future work.

\subsection{Policy Model Training}\label{sec:design:train}

In stage \circled{2}, \sys uses the dataset from stage \circled{1} to perform RL-based post-training via \emph{Generative Reinforcement Policy Optimization} (GRPO) \cite{deepseekr1}. 
The effectiveness of this training heavily depends on the design of the reward function, which must guide the model to generate inputs that explore specific branch conditions.

Existing reward mechanisms—such as human feedback, rule-based pattern matching, or strict execution output comparison—are not ideal for this task. 
Pattern-matching rewards with collected ground-truth answers (\autoref{sec:design:dataset}) are restrictive and harm generalization, as multiple valid inputs can invert a branch. 
Binary execution feedback (e.g., rewarding only on branch reachability) is too sparse for effective training.
To address this, we introduce a coverage distance reward that provides dense, incremental feedback. 
Inspired by directed fuzzing \cite{directed}, our reward function $r(x, y)$ quantifies how closely a generated input $y$'s execution path approaches the target branch, relative to the original input $x$.

Formally, let $T(x)$ and $T(y)$ denote the function-level execution traces of the original and generated inputs, respectively, $F$ as the function containing the target branch, and $C$ ($!C$) as the original (inverted) outcome of the branch condition.
We define their function level \emph{coverage distance} as
\begin{equation}
\label{eq:reward-distance}
\begin{aligned}
d\!\left(x,y\right) 
= \frac{\big\lvert \text{LongestCommonPrefix}(T(x),T(y)) \big\rvert}
       {\lvert T(x)\rvert},
\end{aligned}
\end{equation}

and the reward function $r(x,y)$ as
\begin{equation}
\label{eq:reward}
r(x,y) =
\begin{cases}
d\!\left(x,y\right) & \!\!\!\text{if $F$ not reached by $y$}, \\[4pt]
1   & \!\!\!\!\!\!\!\!\!\!\!\text{if branch reached with $C$ by $y$}, \\[4pt]
2   & \!\!\!\!\!\!\!\!\!\!\!\text{if branch reached with $!C$ by $y$}, \\[4pt]
0.1 & \!\!\!\!\!\!\!\!\!\!\!\text{penalty if $y=x$}. \\[4pt]
\end{cases}
\end{equation}

For example, if the original input's executed function trace towards a branch condition is \lstinline{main->f1->f2->f3:condition1:False}, and the new input's trace is \lstinline{main->f1->f2->f4->f5}, their distance is measured by the number of same functions they executed, resulting in \textbf{\lstinline{3/4=0.75}}.
When the new trace can reach the function containing the target branch but does not invert the condition outcome (\ie{,} has the same effect as the original input), we reward the model by \textbf{\lstinline{1}}.
When the new input successfully inverts the branch outcome, we reward the model by \textbf{\lstinline{2}}.
Notably, since the original input is given and would be rewarded positively by \autoref{eq:reward-distance}, we instruct the model not to respond with the same input and penalize such behavior to avoid \textit{reward hacking} by deducting the reward to \textbf{\lstinline{0.1}}.
As a result, the total reward $r(x,y)$ lies in the range $[0,2]$ before being normalized into GRPO \cite{deepseekr1}.
\subsection{LLM-powered Fuzzing Loop}\label{sec:design:fuzz}
In stage \circled{3} of \sys, we design an LLM-powered fuzzing loop to harness LMs for real-world fuzzing, aiming at enhancing code coverage and vulnerability discovery.
This stage enables a practical and comprehensive evaluation of LMs on the task of fuzzing input generation. 
Because it allows \sys to not only assess accuracy on the generated static dataset but also measure model performance in practical fuzzing campaigns.

\begin{algorithm}[htbp]
\caption{LLM-powered Fuzzing Loop}
\label{alg:fuzz}

\SetKwFunction{FExecuteCheckSave}{ExecuteCheckSave}
\SetKwFunction{FConstructQuestions}{ConstructQuestion}
\SetKwFunction{FFuzzLoop}{FuzzLoop}

\SetKwProg{Fn}{Function}{:}{end}
    $\mathit{Coverage} \gets \varnothing$\;
    $\mathit{Queue} \gets \varnothing$\;
    \While{$\mathsf{Fuzzer.NotStopped()}$}{
        $\mathit{seed} \gets \mathsf{Corpus.ScheduleNext}()$\;
            $\mathit{newSeed} \gets \mathsf{Fuzzer.Mutate}(\mathit{seed})$\;
            $\mathit{feedback} \gets \mathsf{Execute}(\mathit{Program}, \mathit{newSeed})$\;
            $\mathit{hasNewCov} \gets \mathsf{Corpus.CheckSave}(\mathit{newSeed}, \mathit{feedback})$\;
            \If{$\neg\mathit{hasNewCov}$}{
                \textbf{continue};
            }
        
        $\mathit{Coverage} \gets \mathit{Coverage} \cup \mathit{feedback.CoveredBranches}$\;
        
        \ForEach{$\mathit{ub} \in \mathit{feedback.UncoveredBranches}$}{
            \If{$\mathit{ub} \notin \mathit{Coverage}$}{
                $\mathit{question}, \mathit{priority} \gets \FConstructQuestions(\mathit{ub})$\;
                $\mathit{Queue}.\mathsf{Enqueue}(\mathit{question}, \mathit{priority})$\;
            }
        }
    }
    ...


\end{algorithm}

As shown in \autoref{fig:overview}, the fuzzing loop surrounds feeding generated and mutated test inputs into the target programs to reveal potential vulnerabilities through monitoring if the execution crashes.
\autoref{alg:fuzz} describes the process of our fuzzing loop.
It maintains an accumulated \lstinline{Coverage} set of reached branches during fuzzing.
In each iteration, the fuzzer selects a seed from the fuzzing corpus to mutate and execute (lines 3 to 5).
For each mutated seed, the fuzzer examines if it covers previously uncovered code; if it does, the corpus is updated with the new seed and the \lstinline{Coverage} set is updated accordingly; otherwise, the loop continues to the next iteration (lines 7 to 10).

Next, to integrate LMs, \sys extracts uncovered branches for each seed (line 11) and checks if they are \textit{not reached} by previous seeds (line 12).
This check differs from the check in \autoref{alg:dataset} line 14, since in real-world fuzzing, \sys aims at exploring uncovered branches. 
Afterwards, it performs the same \lstinline{ConstructQuestion} process as in \autoref{alg:dataset}.
Notably, instead of querying LMs immediately after constructing each question, we maintain a priority queue (\lstinline{Queue}) to performa better \textit{question scheduling} (line 14).
If an uncovered branch is repeatedly encountered without being reached, it suggests that both LMs and the fuzzing mutator are stuck on that particular branch.
In such cases, \sys uses the negation of the queried count as the \lstinline{priority}, ensuring that LMs explore newly encountered branches instead of repeatedly querying the same question constructed from a challenging branch. 
We omit certain details in the algorithm, including a consumer thread to retrieve questions from the \lstinline{Queue}, querying LMs to execute and save seeds, monitoring crashes, etc.


\section{Implementation}
\label{sec:impl}

We implemented \autoref{alg:dataset} and \autoref{alg:fuzz} in \sys in nearly 5000 lines of Python code.
The training stage is implemented based on the \textit{Verl} \cite{verl} reinforcement learning framework, in which we integrated our coverage distance-based reward calculation.
The reward calculation is based on the source-based code coverage in \textit{LLVM} \cite{llvmsource}, which we modified to collect covered and uncovered branches more efficiently. 
The fuzzing loop (\autoref{alg:fuzz}) is implemented on top of the fuzzer integration in \textit{FuzzBench} \cite{fuzzbench}.

\section{Evaluation}
\label{sec:eval}

In the evaluation of \sys, we aim to answer the following research questions:

\begin{mybullet}

\item \textbf{RQ1:} Can RL-based training enhance the performance of LMs on our fuzzing input generation dataset (\autoref{sec:eval:rl})?

%

\item \textbf{RQ2:} Can trained LMs excel in real-world fuzzing (\autoref{sec:eval:fuzz})?


\item \textbf{RQ3:} How effective are the design choices of \sys (\autoref{sec:eval:abl})?

\end{mybullet}

\PP{Datasets}
To construct our dataset of textual fuzzing input generation (\ie{,} stage \circled{1}), we select 10 real-world targets in total.
They consist of language compilers, interpreters, and database engines, including PHP, CPython, Lua, mruby, NJS, QuickJS, Solidity, SQLite, Sql-parser, and DuckDB.
All of them have manually crafted fuzzing drivers by experts maintained by OSS-Fuzz \cite{ossfuzz}, which offers unified entry points for fair experiments.

To prepare the dataset construction corpus required by \autoref{alg:dataset}, we directly use the initial fuzzing corpus provided by OSS-Fuzz.
For targets that have no provided corpus, we ran a baseline fuzzer (\textit{AFL++} \cite{aflpp}) for 24 hours and collected the fuzzing corpus as our dataset construction corpus.
To limit the dataset size, we randomly sampled up to 1,000 seeds from the initial corpus when its size exceeded this threshold.
As a result, we constructed a dataset with 16338 questions in total for the 10 targets.



\subsection{RL-based Post-training Evaluation}\label{sec:eval:rl}

In this section, we present the results that demonstrate the effectiveness of RL-based post-training on our dataset.

In the training stage (\ie{,} stage \circled{2}), we used \textit{Qwen2.5-7B-Instruct} \cite{qwen} as our base model.
We randomly partitioned the prepared dataset into training and testing sets in a 9:1 ratio and used the training set to train the model using GRPO (\autoref{sec:design:train}) with a batch size of 128. 
To balance stability and exploration, we enabled KL regularization with a low-variance estimator and a coefficient of 0.001. 
Rollouts were performed with eight candidate generations per question under a temperature of 1.0. 
We calculated reward scores on the testing set every two training steps.
\autoref{fig:rl} shows the statistics during training, including the mean of reward scores on the training and testing datasets across 1000 training steps.
The reward scores have a range between 0 to 2 based on \autoref{eq:reward}.
As shown by the trends, both training and testing rewards steadily increased and gradually converged to stable values, indicating that the model successfully learned to align its output with the objective.
Moreover, the testing reward scores closely followed the training scores without significant divergence, suggesting that the model did not overfit to the training set and generalized well to unseen data.
We named the trained model after 1000 steps of RL training as \textit{\textbf{\sys-7B}}.

\begin{figure}[h]
    \centering
    \includegraphics[width=0.82\columnwidth]{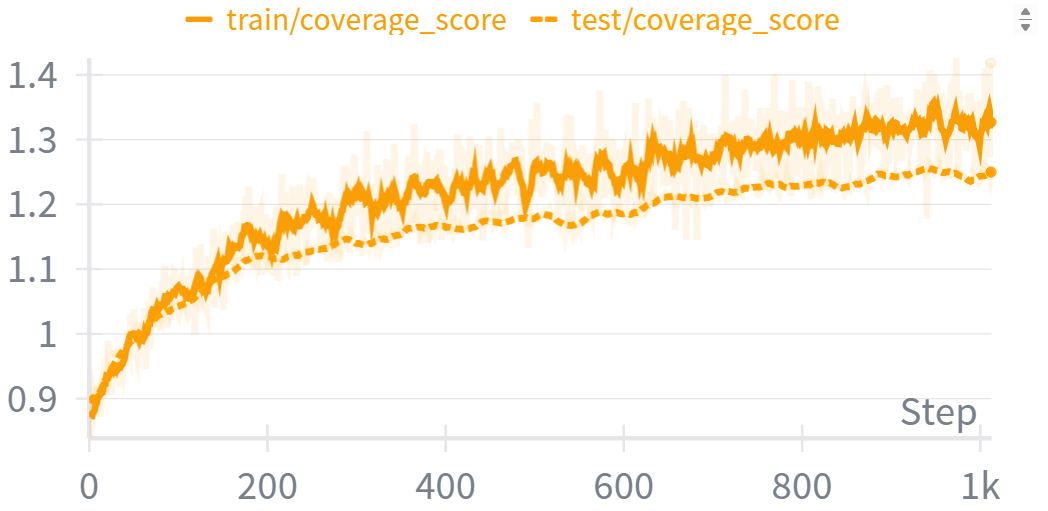}
    \caption{RL-based post-training statistics (reward scores on the training and testing dataset).}
    \label{fig:rl}
    \vspace{-1\baselineskip}
\end{figure}

\begin{table}[h]
\centering
\caption{Static performance of different models on the testing dataset (1640 questions in total)}
\label{tab:static}
\resizebox{\columnwidth}{!}{%
\begin{tabular}{@{}l|ccclc@{}}
\toprule
       & \multicolumn{1}{l}{Qwen2.5-7B} & \multicolumn{1}{l}{Qwen2.5-32B} & \multicolumn{1}{l}{Deepseek-V3} & GPT-o4mini & \multicolumn{1}{l}{\sys-7B} \\ \midrule
Pass@1 & 145 (8\%)                      & 264 (16\%)                      & 391 (23\%)                      & 583 (35\%) & \textbf{830 (50\%)}                        \\
Pass@5 & 342 (20\%)                     & 543 (33\%)                      & 651 (39\%)                      & 811 (49\%) & \textbf{904 (55\%)}                        \\ \bottomrule
\end{tabular}%
}
\end{table}

\autoref{tab:static} shows performance statistics on our dataset from different models.
In addition to \textit{Qwen2.5-7B} and \textit{\sys-7B}, we included both open-source models (\textit{Qwen2.5-32B} \cite{qwen}, \textit{Deepseek-V3} \cite{deepseekr1}) and a closed-source model (\textit{GPT-o4mini} \cite{o4mini}).
Each model was evaluated by answering all questions in the testing dataset, and we measured their accuracy using \textit{Pass@1} and \textit{Pass@5} scores, which capture the fraction of questions correctly answered on the first attempt and within five attempts, respectively.
The results show that \textit{\sys-7B} outperforms both its base model \textit{Qwen2.5-7B} (with nearly 5x higher scores) and larger models (with nearly 60\% higher scores), demonstrating the effectiveness of our RL–based training in boosting reasoning ability on our dataset. 

\begin{table*}[h]
\tiny
\centering
\caption{Fuzzing performance (code region coverage, ratio of correct answers, number of total questions) from different models integrated in the LLM-powered fuzzing loop in \sys (the darker the cell color, the better).}
\label{tab:modelcov}
\resizebox{\textwidth}{!}{%
\begin{tabular}{@{}cccccccccccc@{}}
\toprule
\rowcolor[HTML]{FFFFFF} 
                                                                                      &                                                    & PHP                          & CPython                                              & Lua                          & mruby                        & NJS                                                  & QuickJS                      & Solidity                                             & Sqlite                       & Sql-parser                           & DuckDB                                               \\ \midrule
\rowcolor[HTML]{FFFFFF} 
\multicolumn{1}{c|}{\cellcolor[HTML]{FFFFFF}Baseline}                                 & \multicolumn{1}{c|}{\cellcolor[HTML]{FFFFFF}Cov}   & {\color[HTML]{000000} 82526} & 34470                                                & 7645                         & 7923                         & 8775                                                 & 14377                        & 21032                                                & 26942                        & 1491                                 & 18107                                                \\
\rowcolor[HTML]{E5E5E5} 
\multicolumn{1}{c|}{\cellcolor[HTML]{FFFFFF}}                                         & \multicolumn{1}{c|}{\cellcolor[HTML]{FFFFFF}Cov}   & {\color[HTML]{000000} 82140} & 34870                                                & \cellcolor[HTML]{D2D2D2}7808 & 14036                        & 8393                                                 & 14653                        & 21562                                                & 27707                        & 1951                                 & 36352                                                \\
\rowcolor[HTML]{E5E5E5} 
\multicolumn{1}{c|}{\multirow{-2}{*}{\cellcolor[HTML]{FFFFFF}Qwen2.5-7B}}                 & \multicolumn{1}{c|}{\cellcolor[HTML]{FFFFFF}Ratio} & 0.2\% (7048)                 & 0.1\% (7435)                                         & 0.1\% (7242)                 & 0.1\% (7043)                 & 0.9\% (7352)                                         & 0.5\% (8145)                 & 7\% (7968)                                           & 1.7\% (7142)                 & 1.1\% (6506)                         & 10\% (6417)                                          \\
\rowcolor[HTML]{D2D2D2} 
\multicolumn{1}{c|}{\cellcolor[HTML]{FFFFFF}}                                         & \multicolumn{1}{c|}{\cellcolor[HTML]{FFFFFF}Cov}   & 82298                        & 36364                                                & \cellcolor[HTML]{E5E5E5}7778 & 14486                        & 10840                                                & 14847                        & 21563                                                & 27715                        & 1961                                 & 38180                                                \\
\rowcolor[HTML]{D2D2D2} 
\multicolumn{1}{c|}{\multirow{-2}{*}{\cellcolor[HTML]{FFFFFF}Qwen2.5-32B}}                & \multicolumn{1}{c|}{\cellcolor[HTML]{FFFFFF}Ratio} & 0.3\% (4658)                 & 0.2\% (6929)                                         & 0.2\% (6917)                 & 0.9\% (6771)                 & 2.2\% (6866)                                         & 2.9\% (6160)                 & 18\% (6001)                                          & 5.1\% (6542)                 & 0.9\% (5870)                         & 13\% (6288)                                          \\
\rowcolor[HTML]{BEBEBE} 
\multicolumn{1}{c|}{\cellcolor[HTML]{FFFFFF}}                                         & \multicolumn{1}{c|}{\cellcolor[HTML]{FFFFFF}Cov}   & 83066                        & \cellcolor[HTML]{A7A7A7}36452                        & 7880                         & 14929                        & 11662                                                & 15942                        & \cellcolor[HTML]{A7A7A7}21602                        & 27814                        & 1965                                 & \cellcolor[HTML]{A7A7A7}43769                        \\
\rowcolor[HTML]{BEBEBE} 
\multicolumn{1}{c|}{\multirow{-2}{*}{\cellcolor[HTML]{FFFFFF}Deepseek-V3}}            & \multicolumn{1}{c|}{\cellcolor[HTML]{FFFFFF}Ratio} & 6.8\% (5113)                 & 0.5\% (6210)                                         & 3.4\% (7394)                 & 5.2\% (5518)                 & 7.1\% (5580)                                         & 4.9\% (5709)                 & 19\% (5572)                                          & 10\% (6409)                  & 0.9\% (6388)                         & 18\% (4928)                                          \\
\rowcolor[HTML]{919191} 
\multicolumn{1}{c|}{\cellcolor[HTML]{FFFFFF}}                                         & \multicolumn{1}{c|}{\cellcolor[HTML]{FFFFFF}Cov}   & 88554                        & \cellcolor[HTML]{BEBEBE}34793                        & 8108                         & 15619                        & \cellcolor[HTML]{A7A7A7}13158                        & 17507                        & 21816                                                & 28791                        & 2013                                 & 51070                                                \\
\rowcolor[HTML]{919191} 
\multicolumn{1}{c|}{\multirow{-2}{*}{\cellcolor[HTML]{FFFFFF}GPT-o4mini}}             & \multicolumn{1}{c|}{\cellcolor[HTML]{FFFFFF}Ratio} & 12\% (5662)                  & \cellcolor[HTML]{A7A7A7}0.2\% (6700)                 & 4\% (6953)                   & 9.1\% (6162)                 & \cellcolor[HTML]{A7A7A7}7.9\% (6577)                 & 6\% (6800)                   & 44\% (6451)                                          & 25\% (6615)                  & \cellcolor[HTML]{A7A7A7}1.1\% (6643) & 28\% (5505)                                          \\
\rowcolor[HTML]{A7A7A7} 
\multicolumn{1}{c|}{\cellcolor[HTML]{FFFFFF}}                                         & \multicolumn{1}{c|}{\cellcolor[HTML]{FFFFFF}Cov}   & {\color[HTML]{000000} 83967} & \cellcolor[HTML]{919191}{\color[HTML]{000000} 36966} & {\color[HTML]{000000} 7923}  & {\color[HTML]{000000} 15088} & \cellcolor[HTML]{919191}{\color[HTML]{000000} 14624} & {\color[HTML]{000000} 16941} & \cellcolor[HTML]{BEBEBE}{\color[HTML]{000000} 21585} & {\color[HTML]{000000} 28441} & {\color[HTML]{000000} 1994}          & \cellcolor[HTML]{BEBEBE}{\color[HTML]{000000} 41363} \\
\rowcolor[HTML]{A7A7A7} 
\multicolumn{1}{c|}{\multirow{-2}{*}{\cellcolor[HTML]{FFFFFF}\sys-7B}} & \multicolumn{1}{c|}{\cellcolor[HTML]{FFFFFF}Ratio} & 8.2\% (5772)                 & \cellcolor[HTML]{919191}0.8\% (7736)                 & 2.9\% (7547)                 & 6.9\% (6808)                 & \cellcolor[HTML]{919191}13.5\% (7659)                & 5.1\% (6872)                 & 23\% (6906)                                          & 12\% (6975)                  & \cellcolor[HTML]{919191}1.2\% (6980) & 20\% (5776)                                          \\ \bottomrule
\end{tabular}%
}
\end{table*}

\subsection{Fuzzing Evaluation}\label{sec:eval:fuzz}


In this section, we present the results of evaluating \sys-7B in practical fuzzing campaigns on our selected targets.
We conducted two sets of experiments: the first set compares the performance between different language models in fuzzing (\autoref{sec:eval:model}), and the second set compares our LLM-powered fuzzing loop (stage \circled{3} in \autoref{fig:overview}) with different state-of-the-art fuzzers (\autoref{sec:eval:sota}).
Lastly, we presented the results of newly discovered vulnerabilities in \autoref{sec:eval:bug}.
All the fuzzing experiments lasted for 24 hours for each target, and we ran five trials to calculate the average of code coverage. 
All fuzzing experiments were conducted on a server equipped with an Intel Xeon Gold 5418Y CPU (128 cores) and 512 GB of RAM.

\PP{Data Leakage Prevention} Besides using the initial corpus provided by OSS-Fuzz \cite{ossfuzz} for all fuzzers when starting fuzzing, we augmented this corpus with the “answered seeds” extracted from our constructed training dataset.
As described in \autoref{sec:design:dataset}, each constructed question can be associated with one or more correct answer seeds, which can serve as a form of ground truth. 
By incorporating these answered seeds into the initial corpus, we ensure that the corresponding branches are already covered at the beginning of fuzzing. 
Consequently, questions constructed during training will not reappear during fuzzing, since their associated branches no longer represent uncovered code. 
In other words, all questions encountered by \sys-7B during fuzzing are outside its training set, preventing data leakage and ensuring a fair evaluation of generalization.

\subsubsection{Comparison of Models}\label{sec:eval:model}
We used one of the most widely-used and well-maintained fuzzing engines, \textit{AFL++} \cite{aflpp}, as the base fuzzing engine of stage \circled{3}.
Then, we created different instances of LLM-powered fuzzers named based on different language models they used, such as \textit{AFL++Qwen2.5-7B}, or \textit{AFL++\sys-7B}, etc.
In this subsection, we omit the "\textit{AFL++}" prefix for brevity.

\begin{figure*}
    \centering
    \includegraphics[width=1.0\textwidth]{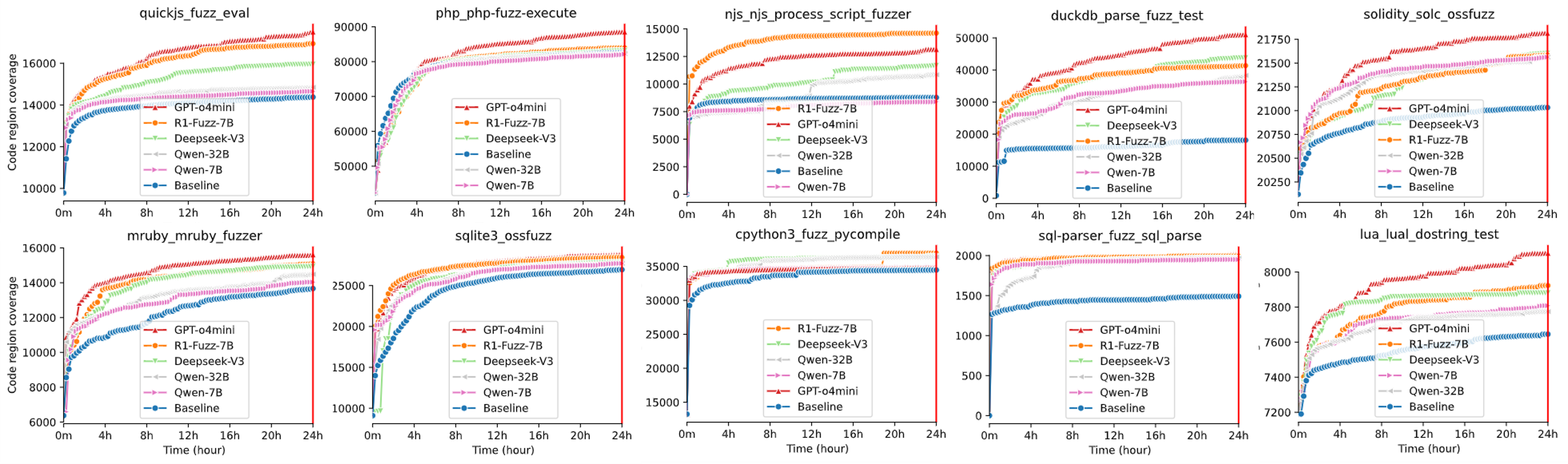}
    \caption{Code Coverage from Different Models}
    \label{fig:modelcov}
\end{figure*}

\autoref{tab:modelcov} shows the results, where the \textit{Baseline} is the original \textit{AFL++} without the LLM-powered loop design.
The \textit{Cov} row indicates the number of covered code regions calculated by the source-based coverage in LLVM \cite{llvmsource}, and the \textit{Ratio} row contains the ratio of correctly answered questions and the total number of asked questions during 24 hours' fuzzing.
For example, for \textit{PHP}, \sys-7B constructed 5772 questions in total in 24 hours, with 473 (8.2\%) input seeds generated by the model that correctly answered the question, \ie{,} reached previously uncovered branches.

The results in the table demonstrate several key findings. 
First, all fuzzers under our LLM-powered fuzzing loops outperform the baseline, confirming the effectiveness of our stage \circled{3} design in \sys. 
On average, the coverage achieved by \sys exceeds the baseline by nearly 40\%, highlighting the substantial gain from incorporating LMs. 
Second, the observed code coverage generally grows in proportion to the ratio of correctly answered questions, \ie{,} models that are more effective at generating correct inputs tend to achieve higher coverage, providing explicit evidence that model reasoning ability translates into tangible fuzzing improvements. 
In a few cases, however, such as target \textit{Sql-parser}, the ratio is higher but the coverage is not, which we attribute to the inherent randomness of fuzzing; notably, the differences remain relatively small.

Comparing across models, we observe a general hierarchy: \textit{Qwen2.5-7B} < \textit{Qwen2.5-32B} < \textit{DeepSeek-V3} < \textit{\sys-7B} < \textit{GPT-o4mini}.
The legends in \autoref{fig:modelcov} are arranged from top to bottom in the order of each model’s coverage performance.
It shows that our RL-based training substantially enhances the base 7B model, enabling it to unlock its latent potential and rival or even surpass much larger models with more parameters.
Moreover, large closed-source models incur costs with ~40 USD per target for 24 hours, while our \sys-7B runs efficiently on local hardware, demonstrating a favorable balance between performance and costs.

For certain special cases, for example, on target Lua, \textit{Qwen2.5-7B} slightly outperforms \textit{Qwen2.5-32B} despite their similar ratios, because the smaller model answered more total questions, underscoring the importance of model efficiency in the intensive fuzzing scenarios. 
For the targets \textit{CPython} and \textit{NJS}, \textit{\sys-7B} even outperforms \textit{GPT-o4mini}, providing further evidence of the necessity of stimulating latent capabilities in low-cost LMs for fuzzing.

Comparing across targets, performance varies for all models: some have higher ratios while others remain less satisfactory.
We attribute this to three possible factors: 
(1) the training dataset can be further improved to better capture real-world fuzzing distributions; 
(2) the question construction may be refined to improve the provided context; 
and (3) limitations may stem from the fundamental ability of the underlying base model, as they exhibit similar weaknesses. 
We leave these directions to future work.

In summary, the LLM-powered fuzzing loop in \sys proves effective in real-world fuzzing, achieving significant gains over the baseline. 
Furthermore, our RL–based post-training enables a cost-efficient 7B model to reach or exceed the performance of larger models. 
These findings demonstrate both the practical effectiveness of our approach and the broader potential of RL training for unlocking the potential of LMs in fuzzing.

\subsubsection{Comparison of State-Of-The-Arts}\label{sec:eval:sota}
Next, we compared the performance of \textit{AFL++\sys-7B} against a general-purpose fuzzer \textit{libFuzzer} \cite{libfuzzer} and several state-of-the-art fuzzers, including grammar-aware fuzzers (\textit{Polyglot} \cite{polyglot}, \textit{Nautilus} \cite{nautilus}, and \textit{Gramatron} \cite{gramatron}) on an available subset of our selected targets.

\autoref{tab:sotacov} presents the number of covered code regions achieved by each approach.
Targets not in \autoref{tab:sotacov} and entries marked \textit{N/A} indicate cases where the open-source implementation of the grammar-based fuzzer does not provide the grammar or specification file required for the corresponding programming language. 
In contrast, our approach builds on the general fuzzer \textit{AFL++} \cite{aflpp} without relying on handcrafted specifications, highlighting the generalization capability of our LLM-powered fuzzing loop.
The results show that \textit{AFL++\sys-7B} consistently outperforms all compared fuzzers.
The average code coverage improvement achieves nearly 75\%, demonstrating the significant effectiveness of integrating LMs into the fuzzing loop (stage \circled{3}). 

\begin{table}[]
\tiny
\centering
\caption{Code region coverage of \sys and other state-of-the-art fuzzers.}
\label{tab:sotacov}
\resizebox{\columnwidth}{!}{%
\begin{tabular}{@{}cccccc@{}}
\toprule
                       & PHP            & mruby          & NJS            & QuickJS        & Solidity       \\ \midrule
libFuzzer              & 58274          & 9738           & 4382           & 12865          & 20673          \\
Polyglot               & 46768          & N/A            & N/A            & 10913          & 20406          \\
Gramatron              & 61943          & 9230           & 7633           & 11550          & N/A            \\
Nautilus               & 57020          & 10102          & 10835          & 13145          & N/A            \\
AFL++\sys-7B & \textbf{83967} & \textbf{15088} & \textbf{14624} & \textbf{16941} & \textbf{21585} \\ \bottomrule
\end{tabular}%
}
\end{table}

\begin{table}[h]
\centering
\caption{Bug numbers}
\label{tab:bugnum}
\resizebox{\columnwidth}{!}{%
\begin{tabular}{@{}cccccccccc@{}}
\toprule
\multicolumn{2}{c}{}                                                                                    & PHP  & Lua & mruby & NJS & QuickJS & Solidity & DuckDB & Total \\ \midrule
\multicolumn{2}{c|}{AFL++}                                                                              & 0    & 0   & 1     & 0   & 0       & 0        & 0      & 1     \\
\multicolumn{2}{c|}{libFuzzer}                                                                          & 0    & 0   & 1     & 0   & 0       & 0        & 0      & 1     \\
\multicolumn{2}{c|}{Nautilus}                                                                           & 0    & N/A & 1     & 0   & 1       & N/A      & N/A    & 2     \\
\multicolumn{2}{c|}{Gramatron}                                                                          & 0    & N/A & 0     & 0   & 0       & N/A      & N/A    & 0     \\
\multicolumn{2}{c|}{Polyglot}                                                                           & 3    & N/A & N/A   & N/A & 0       & 0        & N/A    & 3     \\ \midrule
\multicolumn{1}{c|}{\multirow{5}{*}{\begin{tabular}[c]{@{}c@{}}AFL++\\ \sys\end{tabular}}} & \multicolumn{1}{c|}{Qwen2.5-7B}             & 0    & 1   & 0     & 0   & 0       & 0        & 0      & 1     \\
\multicolumn{1}{c|}{}                                     & \multicolumn{1}{c|}{Qwen2.5-32B}            & 2(2) & 1   & 0     & 0   & 0       & 0        & 2      & 5     \\
\multicolumn{1}{c|}{}                                     & \multicolumn{1}{c|}{Deepseek-V3}            & 2    & 1   & 2     & 0   & 0       & 1        & 3      & 9     \\
\multicolumn{1}{c|}{}                                     & \multicolumn{1}{c|}{GPT-o4mini}             & 4    & 1   & 0     & 2   & 1(1)    & 0        & 7(3)   & 15    \\
\multicolumn{1}{c|}{}                                     & \multicolumn{1}{c|}{\sys-7B} & 6    & 3   & 2     & 2   & 2       & 3        & 4      & 23    \\ \bottomrule
\end{tabular}%
}
\end{table}

\begin{table*}[]
\tiny
\centering
\caption{Ablation study (relative difference of correctly answered ratio and code coverage)}
\label{tab:abl}
\resizebox{\textwidth}{!}{%
\begin{tabular}{@{}cc|cccccccccc@{}}
\toprule
                                                &       & PHP                            & CPython                          & Lua                             & mruby                          & NJS                              & QuickJS                        & Solidity                         & Sqlite                           & Sql-parser                       & DuckDB                           \\ \midrule
\multicolumn{1}{c|}{w/o Trace}                  & Ratio & \cellcolor[HTML]{FFD7D4}-0.5\%                          & \cellcolor[HTML]{FFD7D4}-12.43\% & \cellcolor[HTML]{FFD7D4}-4.17\% & \cellcolor[HTML]{FFD7D4}-2.2\% & \cellcolor[HTML]{FFD7D4}-11.92\% & \cellcolor[HTML]{DCFDDB}0.83\% & \cellcolor[HTML]{FFD7D4}-14.77\% & \cellcolor[HTML]{FFD7D4}-18.06\% & \cellcolor[HTML]{FFD7D4}-14.48\% & \cellcolor[HTML]{FFD7D4}-11.67\% \\ \midrule
\multicolumn{1}{c|}{}                           & Cov   & \cellcolor[HTML]{FFD7D4}-3.1\% & \cellcolor[HTML]{FFD7D4}-2.4\%   & \cellcolor[HTML]{FFD7D4}-3.1\%  & \cellcolor[HTML]{FFD7D4}-6.0\% & \cellcolor[HTML]{FFD7D4}-6.3\%   & \cellcolor[HTML]{FFD7D4}-1.6\% & \cellcolor[HTML]{FFD7D4}-4.3\%   & \cellcolor[HTML]{FFD7D4}-1.7\%   & \cellcolor[HTML]{FFD7D4}-4.1\%   & \cellcolor[HTML]{FFD7D4}-6.2\%   \\
\multicolumn{1}{c|}{\multirow{-2}{*}{w/o Prio}} & Ratio & \cellcolor[HTML]{FFD7D4}-2.7\% & \cellcolor[HTML]{FFD7D4}-0.2\%   & \cellcolor[HTML]{FFD7D4}-5.1\%  & \cellcolor[HTML]{FFD7D4}-2.6\% & \cellcolor[HTML]{FFD7D4}-1.1\%   & \cellcolor[HTML]{FFD7D4}-1.4\% & \cellcolor[HTML]{FFD7D4}-10.3\%  & \cellcolor[HTML]{FFD7D4}-1.3\%   & \cellcolor[HTML]{FFD7D4}-7.4\%   & \cellcolor[HTML]{FFD7D4}-5.3\%   \\ \bottomrule
\end{tabular}%
}
\end{table*}

\begin{table}[h]
\centering
\caption{Bug details}
\label{tab:bug}
\resizebox{\columnwidth}{!}{%
\begin{tabular}{@{}cccc@{}}
\toprule
   & Target   & Status        & Link                                                  \\ \midrule
1  & PHP      & Fixed         & https://github.com/php/php-src/issues/18845           \\
2  & PHP      & Duplicated & https://github.com/php/php-src/issues/18844           \\
3  & PHP      & Fixed         & https://github.com/php/php-src/issues/18838           \\
4  & PHP      & Fixed         & https://github.com/php/php-src/issues/19303           \\
5  & PHP      & Fixed         & https://github.com/php/php-src/issues/19304           \\
6  & PHP      & Fixed         & https://github.com/php/php-src/issues/19305           \\
7  & PHP      & Fixed         & https://github.com/php/php-src/issues/19306           \\
8  & PHP      & Fixed         & https://github.com/php/php-src/issues/19844           \\
9  & NJS      & Fixed         & https://github.com/nginx/njs/issues/918               \\
10 & NJS      & Fixed         & https://github.com/nginx/njs/issues/921               \\
11 & QuickJS  & Fixed         & https://github.com/bellard/quickjs/issues/412         \\
12 & QuickJS  & Fixed      & https://github.com/bellard/quickjs/issues/441         \\
13 & QuickJS  & Duplicated & https://github.com/bellard/quickjs/issues/413         \\
14 & mruby    & Fixed         & N/A                                                   \\
15 & mruby    & Confirmed     & https://github.com/mruby/mruby/issues/6584            \\
16 & Lua      & Fixed         & https://groups.google.com/g/lua-l/c/IJ30td8P9EY       \\
17 & Lua      & Confirmed     & https://github.com/ligurio/lua-c-api-tests/issues/132 \\
18 & Lua      & Confirmed      & https://github.com/ligurio/lua-c-api-tests/issues/155 \\
19 & DuckDB   & Fixed         & https://github.com/duckdb/duckdb/issues/17781         \\
20 & DuckDB   & Confirmed     & https://github.com/duckdb/duckdb/issues/17780         \\
21 & DuckDB   & Confirmed     & https://github.com/duckdb/duckdb/issues/17734         \\
22 & DuckDB   & Confirmed     & https://github.com/duckdb/duckdb/issues/18448         \\
23 & DuckDB   & Confirmed     & https://github.com/duckdb/duckdb/issues/18449         \\
24 & DuckDB   & Confirmed     & https://github.com/duckdb/duckdb/issues/18450         \\
25 & DuckDB   & Confirmed     & https://github.com/duckdb/duckdb/issues/18451         \\
26 & DuckDB   & Confirmed     & https://github.com/duckdb/duckdb/issues/18452         \\
27 & Solidity & Reported      & https://github.com/ethereum/solidity/issues/16071     \\
28 & Solidity & Reported      & https://github.com/ethereum/solidity/issues/16069     \\
29 & Solidity & Reported      & https://github.com/argotorg/solidity/issues/16202     \\ \bottomrule
\end{tabular}%
}
\end{table}

\subsubsection{Discovery of Vulnerabilities}\label{sec:eval:bug}

We further evaluate the effectiveness of \sys fuzzers (different models integrated with \textit{AFL++} \cite{aflpp}) in vulnerability discovery. 
We ran all fuzzers on the latest version of our selected targets to discover previously unknown bugs.
The results are shown in \autoref{tab:bugnum}. 
The numbers in parentheses indicate unique vulnerabilities discovered by other fuzzers or models but not by \textit{\sys-7B}.
Traditional general fuzzers, including \textit{AFL++} and \textit{libFuzzer}, discovered two bugs, and state-of-the-art grammar-aware fuzzers (\textit{Nautilus}, \textit{Gramatron}, \textit{Polyglot}) found five bugs in total after manual analysis. 
In contrast, our LLM-powered fuzzing loop in \sys outperformed existing methods, uncovering 29 unique vulnerabilities that are previously unknown.
These vulnerabilities span diverse projects, including PHP, Lua, mruby, NJS, QuickJS, Solidity, and DuckDB, with 24 confirmed or fixed by developers shown in \autoref{tab:bug}. 
Compared with the non-LLM fuzzer, \sys revealed over 5× more vulnerabilities, highlighting the practical advantage of integrating language models into fuzzing.

Among different models in our LLM-powered fuzzing loop, our post-trained model, \textit{\sys-7B}, outperforms both smaller and larger non-specialized models.
While a few bugs remain exclusive to other models, the majority of vulnerabilities are revealed by our model.
The comparison with untuned smaller models emphasizes this observation: \textit{Qwen2.5-7B} and \textit{Qwen2.5-32B} identified only 1 and 5 bugs, respectively, whereas \textit{\sys-7B} uncovered a total of 23 bugs.
It demonstrates that our RL-based training effectively equips a small model with specialized ability on the target project.
When combined with the efficiency of the small model, this approach proves highly effective for intensive fuzzing tasks.

In summary, the vulnerability discovery results validate the effectiveness of our LLM-powered fuzzing loop and the necessity of RL-based post-training.

\subsection{Ablation Study}\label{sec:eval:abl}

In this section, we conducted ablation experiments to assess the effectiveness of key design choices in \sys, including the \textit{coverage-slicing-based question construction} and the \textit{question scheduling} design in our LLM-powered fuzzing loop.

\subsubsection{Question Format}

We first investigated the impact of function-trace extraction in the question construction process. 
Specifically, we considered an ablation variant (\textit{w/o Trace}) in which the \lstinline{ConstructQuestion} function (see \autoref{alg:dataset} and \autoref{alg:fuzz}) was modified to exclude the call-stack traces of the target branch. 
In this setting, the constructed question body contained only the code of the immediate function associated with the uncovered branch, without additional contextual information from caller functions.
To obtain an accurate comparison, we did not implement this variant as a separate fuzzing instance. Instead, within the same fuzzer, both the original question and the \textit{w/o Trace} question were constructed for each branch and issued to the model.
This design ensures that both formats are evaluated on the identical set of questions under realistic fuzzing conditions. 
We then calculated the \textit{answer ratio}, \ie{,} the fraction of correctly answered questions over the total number of questions, and reported the relative difference between the two variants.
The results are presented in \autoref{tab:abl}, which demonstrates that the absence of function-trace information generally leads to degraded performance by 8\% on correct answer ratios. 
Across most targets, the reduction in the answer rate confirms that providing the model with execution context—beyond the local function code—is essential for enabling effective reasoning about how to reach uncovered branches.

Besides supplying a broader execution context, we regard further exploration of more question construction strategies, which adaptively balance question validity with contextual completeness, as a valuable future direction.

\subsubsection{Question Scheduling}
We further investigated the effect of our question scheduling mechanism, which determines how constructed questions are prioritized for LM queries during fuzzing. 
As outlined in \autoref{sec:design:fuzz}, \sys continuously extracts uncovered branches to construct questions during fuzzing.
Instead of immediately querying the LMs, \sys enqueues the question into a priority queue for scheduling.
We compared two variants: a naive strategy that queries each constructed question immediately (\textit{w/o Prio}), and the default scheduling mechanism in \sys, which assigns priorities based on the negation of the queried count. 
In the latter, branches that remain unreached despite repeated attempts are gradually deprioritized, thereby encouraging exploration of newly encountered branches.
We implemented the variants into two fuzzing instances.
\autoref{tab:abl} presents their relative differences in code coverage and answer ratio.
The result shows that priority-based scheduling consistently achieves higher answer ratios and successfully leads to higher code coverage by the differences of 3.5\% and 3.8\%, respectively. 
This indicates that our scheduling mechanism can improve query efficiency by avoiding redundant attempts to further enhance the diversity of explored paths.

%

\section{Related Work}
\label{sec:rel}

\subsection{LLM for Fuzzing}

Since the prevalence of LLMs, there are many related works applying LLMs to the fuzzing area, including fuzzing input generation \cite{fuzz4all, fuzzbusybox, clozemaster, codamosa, covrlfuzz, llmif, meng2024large}, fuzzing driver or unit test code generation \cite{promptfuzz, 10.1145/3597926.3598067, coverup, rug}, and generator or mutator code generation \cite{metamut, zhang2025low}.
For example, PromptFuzz \cite{promptfuzz} uses LLMs to analyze library source code for generating and evolving fuzzing driver code that harnesses library APIs to test them.
MetaMut \cite{metamut} harnesses LLMs to generate grammar-aware mutator code to test compilers for the C language.
Similarly, G2Fuzz \cite{zhang2025low} harnesses LLMs to write input generator code for certain formats, such as PDF, ELF, and PNG, to test corresponding processors.
For fuzzing input generation, Fuzz4All \cite{fuzz4all} uses LLMs to read documentation to generate feature-guided inputs.
Asmita et al.~\cite{fuzzbusybox} leverages LLMs to generate inputs for BusyBox with crash-reuse heuristics.
Clozemaster \cite{clozemaster} and CovRL-Fuzz \cite{covrlfuzz} use LLMs to infill or mutate partial fuzzing input for Rust and JavaScript, respectively.
Compared to them, which depend on extra documentation or the input seed itself, \sys is the first framework to post-train small LMs to effectively reason about deep program source code for fuzzing input generation.



\subsection{LLM Post-training}

Recent works have increasingly applied reinforcement learning (RL) as a post-training paradigm to enhance the reasoning, alignment, and domain specialization of large language models. 
DeepSeek-R1 \cite{deepseekr1} highlights the potential of pure RL (by GRPO) in enabling LLMs to reason and benefit from chains of thought, revealing both strengths in problem-solving and challenges in controllability and safety. 
Compiler-R1 \cite{compilerr1} uses RL for compiler auto-tuning, achieving reductions in intermediate representation instruction counts. 
Memory-R1 \cite{memoryr1} equips LLMs with RL-trained agents for adaptive memory management and utilization. 
SWE-RL \cite{swerl} applies RL-based reasoning to real-world software engineering by training on massive software evolution data, enabling a medium-scale model to achieve state-of-the-art performance on SWE-bench while generalizing to out-of-domain reasoning tasks. 
RLSF \cite{rlsf} leverages symbolic feedback from solvers and provers as fine-grained RL signals, enabling smaller models to surpass larger ones on program synthesis, chemistry, and logical reasoning benchmarks.  
Compared to them, \sys is the first framework that demonstrates RL-based post-training is effective on practical fuzzing input generation beyond static benchmark and finds real-world vulnerabilities that are previously unknown.



\section{Conclusion}
\label{sec:conclusion}

In this paper, we introduced \sys, a reinforcement learning-based framework for post-training language models to enhance complex textual fuzzing. 
\sys addresses the key challenges of leveraging LMs for testing complex, constraint-rich software by introducing two core techniques: coverage-slicing-based question construction, which systematically decomposes deep program logic into reasonable questions, and a distance-based reward mechanism that provides fine-grained feedback during RL-based post-training. 
By specializing a cost-efficient 7B-parameter model rather than relying on expensive larger models, \sys significantly improves accessibility and scalability while maintaining competitive fuzzing performance. 
Our evaluation demonstrates that \sys not only achieves higher coverage—up to 75\% more than state-of-the-art fuzzers—but also proves highly practical by uncovering 29 previously unknown vulnerabilities in real-world projects. 
These results highlight the potential of lightweight, specialized LMs to reason about program semantics and advance the state of the art in fuzzing sophisticated textual targets.

\balance
\bibliographystyle{ACM-Reference-Format}

\bibliography{p,conf}
\end{document}